\documentclass[aps,prl,twocolumn,superscriptaddress,showpacs]{revtex4}

\usepackage{graphicx}

\newcommand\pictc[5]{\begin{figure}[t]
                       \centerline{\vspace{-2mm}
\includegraphics[width=#1\columnwidth,height=0.7\textheight,keepaspectratio]{#3}}
                \protect\caption{\protect\label{fig:#4} #5}\vspace{-2mm}
                    \end{figure}            }
\newcommand\pict[4][1]{\pictc{#1}{!tb}{#2}{#3}{#4}}

\newcommand\pictwc[5]{\begin{figure*}[t]
                       \centerline{\vspace{-2mm}
                       \includegraphics[width=#1\textwidth]{#3}}
                \protect\caption{\protect\label{fig:#4} #5}\vspace{-2mm}
                    \end{figure*}            }
\newcommand\pictw[4][1]{\pictwc{#1}{!tb}{#2}{#3}{#4}}

\newcommand\rpict[1]{\ref{fig:#1}}

\newcommand\leqt[1]{\protect\label{eq:#1}}

\newcounter{Fig}

\begin{document}

\title{Photonic Bloch oscillations and Zener tunneling in two-dimensional optical lattices}

\author{Henrike Trompeter}

\affiliation{Nonlinear Physics Centre and Laser Physics Center,
Centre for Ultra-high bandwidth Devices for Optical Systems (CUDOS),
Research School of Physical Sciences and Engineering,
Australian National University, Canberra ACT 0200, Australia}

\affiliation{Institute of Condensed Matter Theory and Solid State Optics,
Friedrich-Schiller-University Jena, Max-Wien-Platz 1, 07743 Jena, Germany}

\author{Wieslaw Krolikowski}
\author{Dragomir N. Neshev}
\author{Anton S. Desyatnikov}
\author{Andrey A. Sukhorukov}
\author{Yuri S. Kivshar}

\affiliation{Nonlinear Physics Centre and Laser Physics Center,
Centre for Ultra-high bandwidth Devices for Optical Systems (CUDOS),
Research School of Physical Sciences and Engineering,
Australian National University, Canberra ACT 0200, Australia}

\author{Thomas Pertsch}
\author{Ulf Peschel}
\author{Falk Lederer}

\affiliation{Institute of Condensed Matter Theory and Solid State Optics,
Friedrich-Schiller-University Jena, Max-Wien-Platz 1, 07743 Jena, Germany}

\begin{abstract}
We report on the first experimental observation of photonic Bloch oscillations and Zener tunneling in two-dimensional periodic systems. We study the propagation of an optical beam in a square photonic lattice superimposed on a refractive index ramp, and demonstrate the tunneling of light from the first to the higher-order transmission bands of the lattice bandgap spectrum, associated with the spectral dynamics inside the first Brillouin zone and corresponding oscillations of the primary beam.
\end{abstract}

\pacs{42.82.Et, 42.25.Bs, 42.65.Wi}

\maketitle

One of the fundamental concepts of quantum mechanics is a duality of particles and waves. Electromagnetic waves propagating in a periodic dielectric medium can behave in a rather similar fashion to electrons in a crystalline potential. Hence many effects originally predicted in solid-state physics can be observed by monitoring light propagation in photonic structures.

Two well-known fundamental phenomena associated with the propagation of waves and quantum particles in periodic media under the action of an external driving force are {\em Bloch oscillations}~\cite{Bloch:1928-555:RAR} and {\em Zener tunneling}~\cite{Zener:1934-523:RAR}. Indeed, in a periodic potential the dynamics of a particle is dramatically affected by the bandgap structure of its energy spectrum. Moreover, due to the existence of the maximum wavenumber determined by the edge of the Brillouin zone, even particles associated with a single spectral band can behave in a rather unexpected way. For example, they do not follow the direction of a driving force, but instead perform an oscillatory motion, the so-called Bloch oscillations. Such oscillations occur because the presence of an external force causes the particle to gain momentum and approach the Bragg resonance. Since the band structure is periodic, the initial field distribution is recovered after one crossing of the Brillouin zone. Therefore, one observes an oscillatory motion but no net shift of the particle in a real space.

In order to explain why electrons move in a crystalline lattice under the action of a dc electric field, this single-band picture of the Bloch oscillations should be extended to take into account coupling to other bands, the effect known as Zener tunneling~\cite{Zener:1934-523:RAR}. If the variation of the superimposed linear potential within a unit cell is comparable with the size of the gap to the adjacent band, an intra-band transition occurs. The strongest Zener tunneling takes place when the particle reaches the edge of the first Brillouin zone where the gap is smallest.

While originally predicted in the context of electrons in crystals, Bloch oscillations and Zener tunneling were also extensively investigated in different physical systems, including electrons in semiconductor superlattices~\cite{Feldmann:1992-7252:PRB, Ghulinyan:2005-127401:PRL} and cold atoms in optical lattices~\cite{Dahan:1996-4508:PRL, Anderson:1998-1686:SCI}.
Recent progress in the fabrication and investigation of complex optical nanostructures has allowed for direct experimental observations of many phenomena related to the wave propagation and interference. One-dimensional optical Bloch oscillations were observed in dielectric structures with a transversely superimposed linear ramp of the refractive index~\cite{Pertsch:1999-4752:PRL, Morandotti:1999-4756:PRL, Sapienza:2003-263902:PRL, Agarwal:2004-97401:PRL}. A periodic distribution of the refractive index plays a role of the crystalline potential, and the index gradient acts similar to an external force in a quantum system. It causes the beam to move across the structure where it experiences Bragg reflection on the high-index and total internal reflection on the low-index side of the structure, resulting in an optical analogue of Bloch oscillations. These experiments have recently been followed by the observations of tunneling in one-dimensional waveguide arrays~\cite{Trompeter:2005-QFA2:ProcCLEO, Usievich:2004-841:OPSR}. It was demonstrated that for a small spectral gap even with a moderate strength of the index gradient, reasonably high transmission rates from the first to the higher-order bands can be achieved facilitating a direct observation of the optical Zener tunneling.

\pict[0.9]{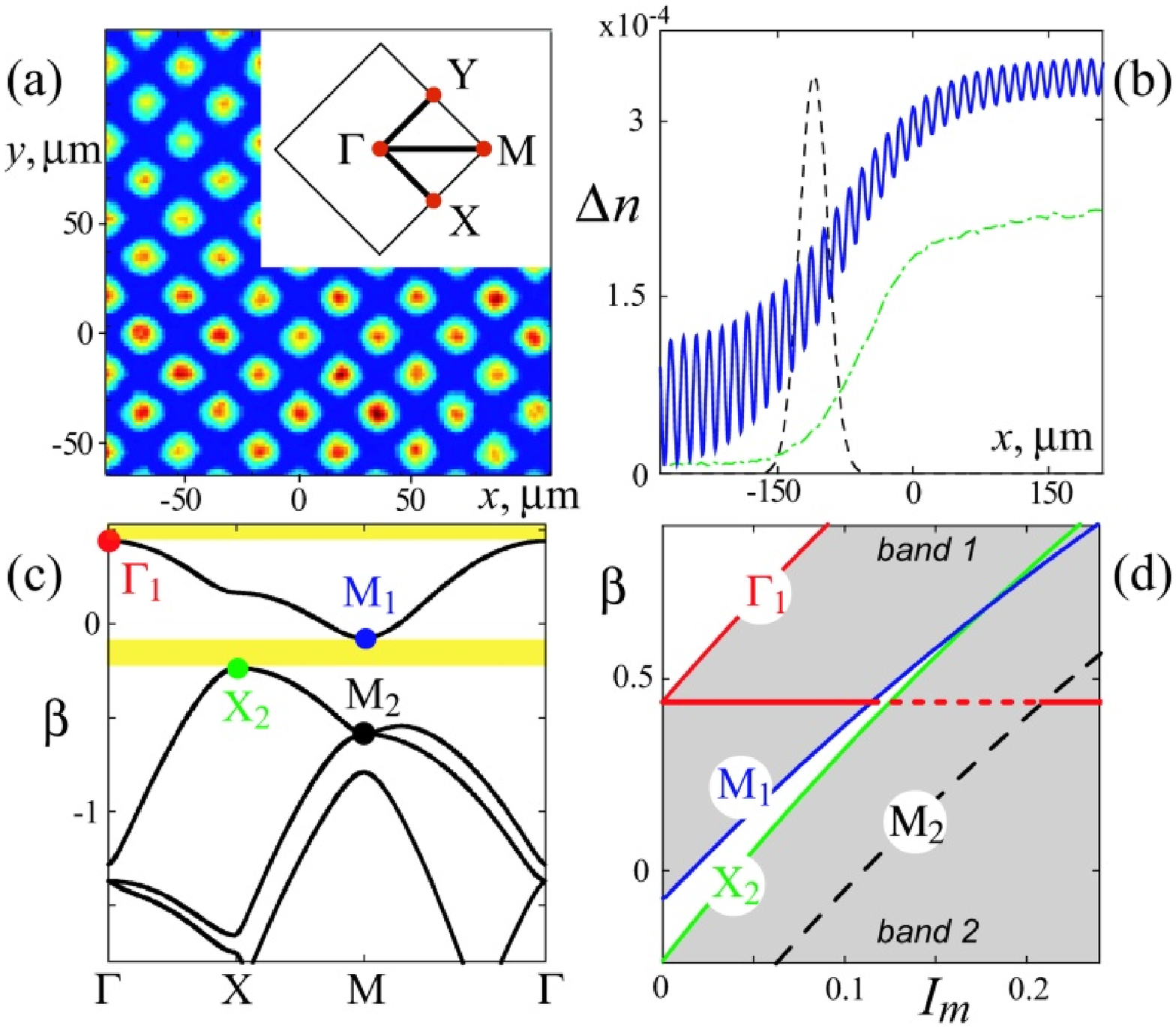}{bandgap}{
(color online) (a) Interference pattern of four beams forming a square lattice; the inset shows symmetry points in the first Brillouin zone. (b) Experimentally measured profile of the background illumination $I_m$ (dash-dotted), and corresponding change in the refractive index (solid). Dashed line indicates the initial position of the probe beam. (c) Calculated band structure of the lattice for $I_m=0$.
(d) Band structure of the lattice vs. $I_m$. Horizontal line is the adiabatic value of propagation constant; the dashed segment indicates Zener tunneling through the gap.}

All direct experimental observations of Bloch oscillations and Zener tunneling performed so far were limited to one-dimensional geometries. 
New effects may be associated with these phenomena in systems of higher dimensionality~\cite{Kolovsky:2003-63601:PRA, Witthaut:2004-41:NJP}.
In a two-dimensional periodic potential the wave follows complex Lissajous-type trajectories when the direction of the static force does not coincide with a principal axis of the lattice. Additionally, the process of Zener tunneling becomes nontrivial as the band-gap structure can cause an enhanced tunneling in preferred directions determined by the lattice symmetries.

In this Letter, we report on the first experimental observation of Bloch oscillations and Zener tunneling in two-dimensional (2D) periodic systems. By launching a laser beam into an optically-induced lattice with a superimposed index gradient, we observe the occurrence of 2D Bloch oscillations and tunneling of light into the second transmission band of the lattice.

The lattice is created by interfering four mutually coherent ordinary-polarized broad beams in a biased photorefractive crystal~\cite{Efremidis:2002-46602:PRE, Fleischer:2003-147:NAT, Neshev:2003-710:OL}. The periodic light intensity distribution inside the crystal has the form of a square lattice shown in Fig.~\rpict{bandgap}(a), described by the expression
\begin{equation} \leqt{I_g}
   I_g(x,y) = A^2 \cos^2(\pi X/d) \cos^2(\pi Y/d),
\end{equation}
where $(X,Y)\equiv(x \mp y)/\sqrt 2$, $x$ and $y$ are the transverse coordinates, $A$ is the normalized amplitude, and $d$ is the lattice period. 
Such a periodic light pattern induces a 2D modulation of the refractive index for the extraordinary polarized probe beam. The ordinary polarized lattice beams, however, remain stationarity along the whole length of the crystal~\cite{Efremidis:2002-46602:PRE, Fleischer:2003-147:NAT}. Orientation of the lattice is set to $45^\circ$ with respect to the $c$-axis of the crystal (horizontal in all figures; coincides with $x$ axis in our notation) in order to reduce the effect of the intrinsic anisotropy of the photorefractive nonlinearity~\cite{Desyatnikov:2005-869:OL}. In order to create a transverse refractive index gradient we illuminate the crystal from the top with an incoherent white light which is modulated transversely, but is  constant along the crystal length. The measured transverse intensity profile of the white light is shown in Fig.~\rpict{bandgap}(b) (dash-dotted line) and it can be well approximated as
\begin{equation} \leqt{index}
   I_m(x) = B \left[ 1 + \tanh \left( x / \eta \right) \right] / 2 ,
\end{equation}
where the parameter $\eta$ determines the extent of the induced index ramp. The total induced refractive index pattern is then given by [Fig.~\rpict{bandgap}(b), solid]
\begin{equation} \leqt{dn}
   \Delta n(x,y)= \gamma \frac{I_g(x,y)+I_m(x)}{1+I_g(x,y)+I_m(x)},
\end{equation}
where $\gamma$ can be tuned by varying the bias voltage.
Then, we simulate the propagation of a probe optical beam by solving paraxial equation for the normalized electric field envelope $E(x,y,z)$,
\begin{equation} \leqt{nls}
   i \frac{\partial E}{\partial z} 
   + \frac{\lambda}{4 \pi n_0}
        \left(   \frac{\partial^2 E}{\partial x^2}
               + \frac{\partial^2 E}{\partial y^2} \right) 
   + \frac{2\pi}{\lambda} \Delta n(x,y) E 
   = 0,
\end{equation}
where $z$ is the propagation distance. We use the following parameters to match our experimental conditions: wavelength in vacuum $\lambda=0.532~\mu$m, linear refractive index of the medium $n_0 = 2.35$, lattice period $d=20\mu$m and amplitude $A=0.5$, $\gamma=6.5$, $\eta=100\mu$m, and $B=1$.

\pict[0.9]{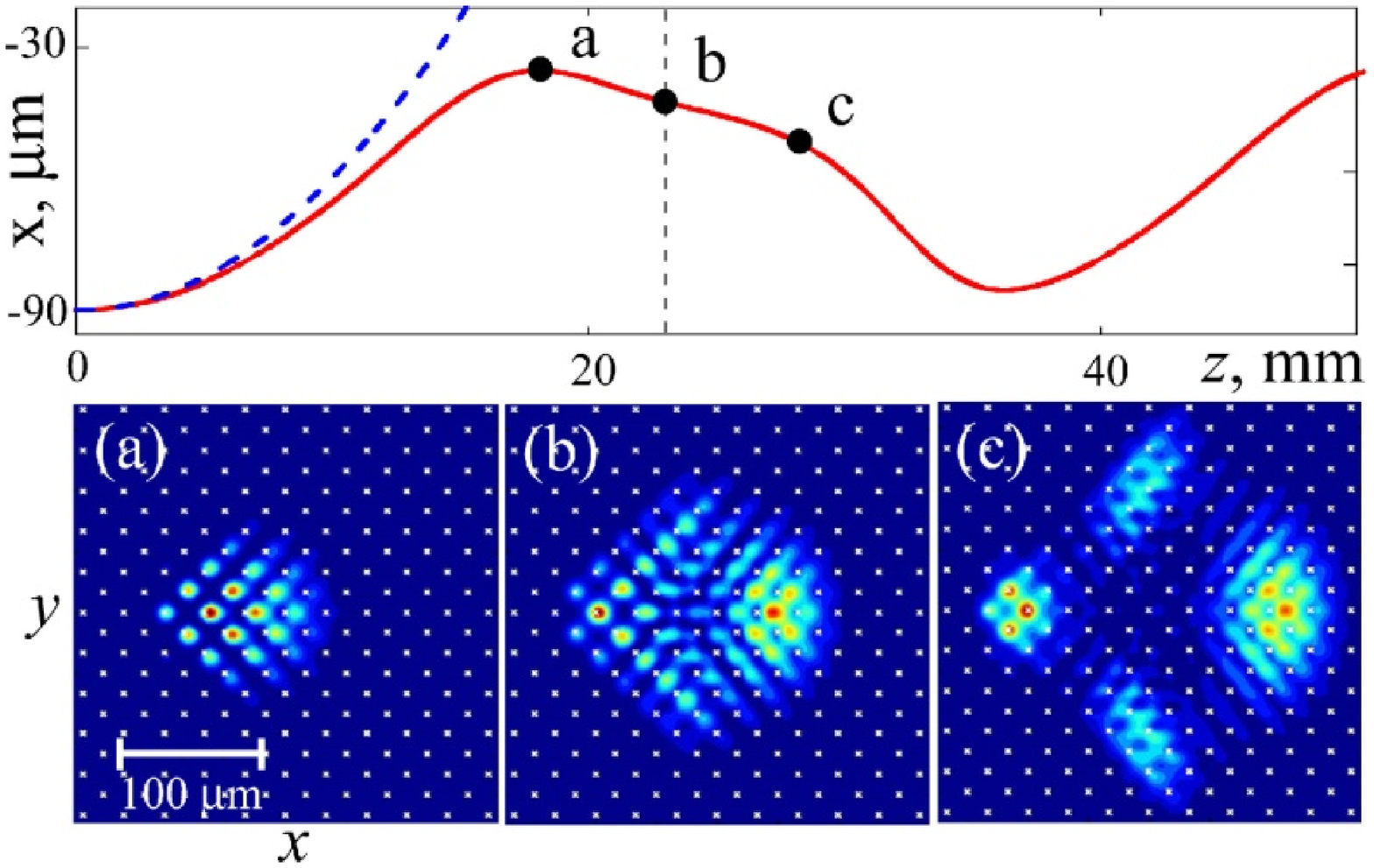}{numerics}{
(color online) (top) Numerically calculated position of the beam's center of mass; solid line -- Bloch oscillations on the lattice; dashed line -- beam deflection without the lattice. The vertical line indicates the position of the exit facet of the crystal. The light intensity distributions corresponding to the positions (a), (b), and (c) are depicted in the three snapshots.}
\pictw[0.8]{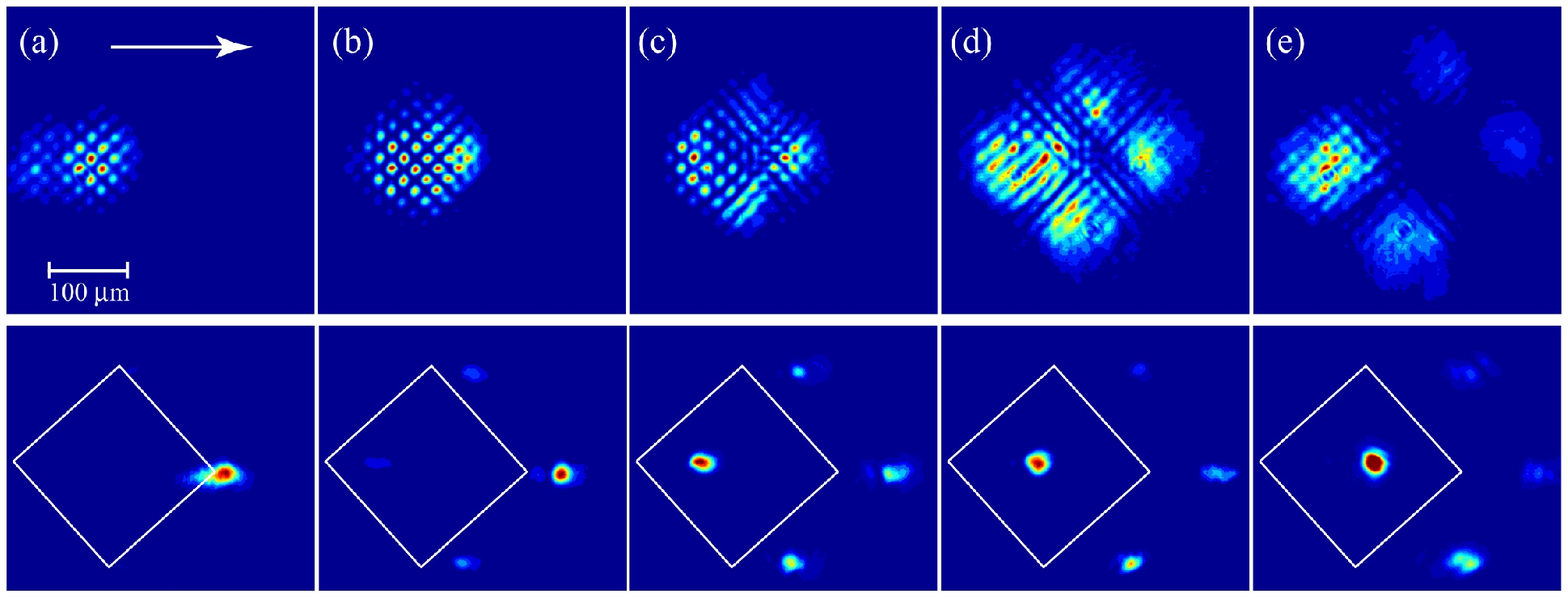}{experiment}{
(color online) Light intensity profiles (top row) and corresponding Fourier spectra (bottom) measured at the crystal output for different initial inclinations of the input beam. The inclination is (a) $-0.095^\circ$, (b) $-0.035^\circ$, (c) $0^\circ$, (d) $0.035^\circ$, and (d) $0.12^\circ$. The arrow in (a) indicates the direction of the index gradient. White diamonds indicate the first Brillouin zone.}

Propagation of waves in a homogeneous lattice (for a fixed $I_m$) is characterized by dispersion of Bloch waves which profiles have the form $E = \Psi(X,Y) \exp(i K_X X + i K_Y Y + i \beta z)$, where ${\bf K}=(K_X,K_Y)$ is the Bloch wavevector, the eigenvalue $\beta$ is the propagation constant, and eigenmode profile $\Psi$ has the same periodicity as the underlying lattice. We show the first Brillouin zone in the inset of Fig.~\rpict{bandgap}(a), and the dispersion curves characterizing the band-gap spectrum are presented in Fig.~\rpict{bandgap}(c) for $I_m=0$, where the high-symmetry points are marked for the first two bands.
The background illumination of the crystal, $I_m(x)$, results in the index distribution which represents a periodic structure imposed onto a monotonically increasing background. Because of the saturable character of the photorefractive nonlinearity, the contrast of the resulting index grating varies transversely across the crystal, and the complete gap decreases, finally closing for high background intensities, see Fig.~\rpict{bandgap}(d). Experiments were performed in the area of largest gradient where the gap is still open.

We now model the propagation of an input Gaussian beam with a full width at half intensity maximum of 37.5~$\mu$m launched straight along the lattice, and plot in Fig.~\rpict{numerics}(top) the evolution of the beam center of mass, defined as $\int x |E|^2 d{\bf r}/\int |E|^2 d{\bf r}$. As expected, the beam oscillates but, in contrast to the canonical Bloch-oscillations~\cite{Kolovsky:2003-63601:PRA}, the trajectories are deformed due to the transverse variations of the depth of the periodic potential. According to the adiabatic theory, the Zener tunneling from the initial $\Gamma_1$ point in Figs.~\rpict{bandgap}(c,d) occurs when the effective propagation constant $\beta$ reaches the gap edge M$_1$ and then {\em tunnels through the gap} to the point M$_2$, see the horizontal line in Fig.~\rpict{bandgap}(d). Due to the square symmetry of the lattice this process results in the splitting of the initial beam into four parts, as shown in Figs.~\rpict{numerics}(a-c). The intensity maxima of the strongest (reflected) portion are centered on the high-index points of the lattice. Hence this beam belongs to the fundamental band, and it keeps propagating in an oscillatory fashion. The other three beams are formed as a result of the tunneling to the second band, which is confirmed by the fact that their maxima are located in between the index maxima of the lattice. 
Weak or even vanishing gap on the high-index side promotes strong tunneling of light even for a moderate strength of the index gradient.

We study experimentally the oscillations and tunneling in 2D optical lattices using a 23~mm long, strontium barium niobate (SBN:65) photorefractive crystal, biased with an electric field of 5000~V/cm. A light from a green cw laser at 532~nm is used to create the lattice and the probe beam, while additional while light illuminator is used to induce the transverse index gradient. First, we test the effect only of the induced index gradient on the beam propagation by switching off the lattice and launching the extraordinary polarized signal beam into the region with the steepest gradient. Measurements of the field at the output facet of the crystal performed with a CCD camera showed that the beam experiences a maximum transverse shift of 350~$\mu$m. In the next step, we launch the signal beam in the presence of optical lattice and the index gradient. Respective images of the signal beam at the exit facet of the crystal are displayed in Figs.~\rpict{experiment}(a-e, top row). It is not possible to follow the evolution of the beam inside the crystal, but we can infer the details of its behavior by varying the incident angle of the probe beam. For angles below the Bragg resonance, the excitation by the probe with different transverse wavenumber is equivalent to different starting points in the Brillouin zone. Therefore, when we launch the beam at different input angles we can scan the field evolution through the Bloch oscillation at the output facet of the crystal. In this way, we are able to monitor different stages of the Bloch oscillations, which would normally require crystals of different lengths. In Fig.~\rpict{experiment}(a-e, top row) we can clearly see the predicted reshaping of the beam and the tunneling of the beam energy into three different channels.

\pict[0.9]{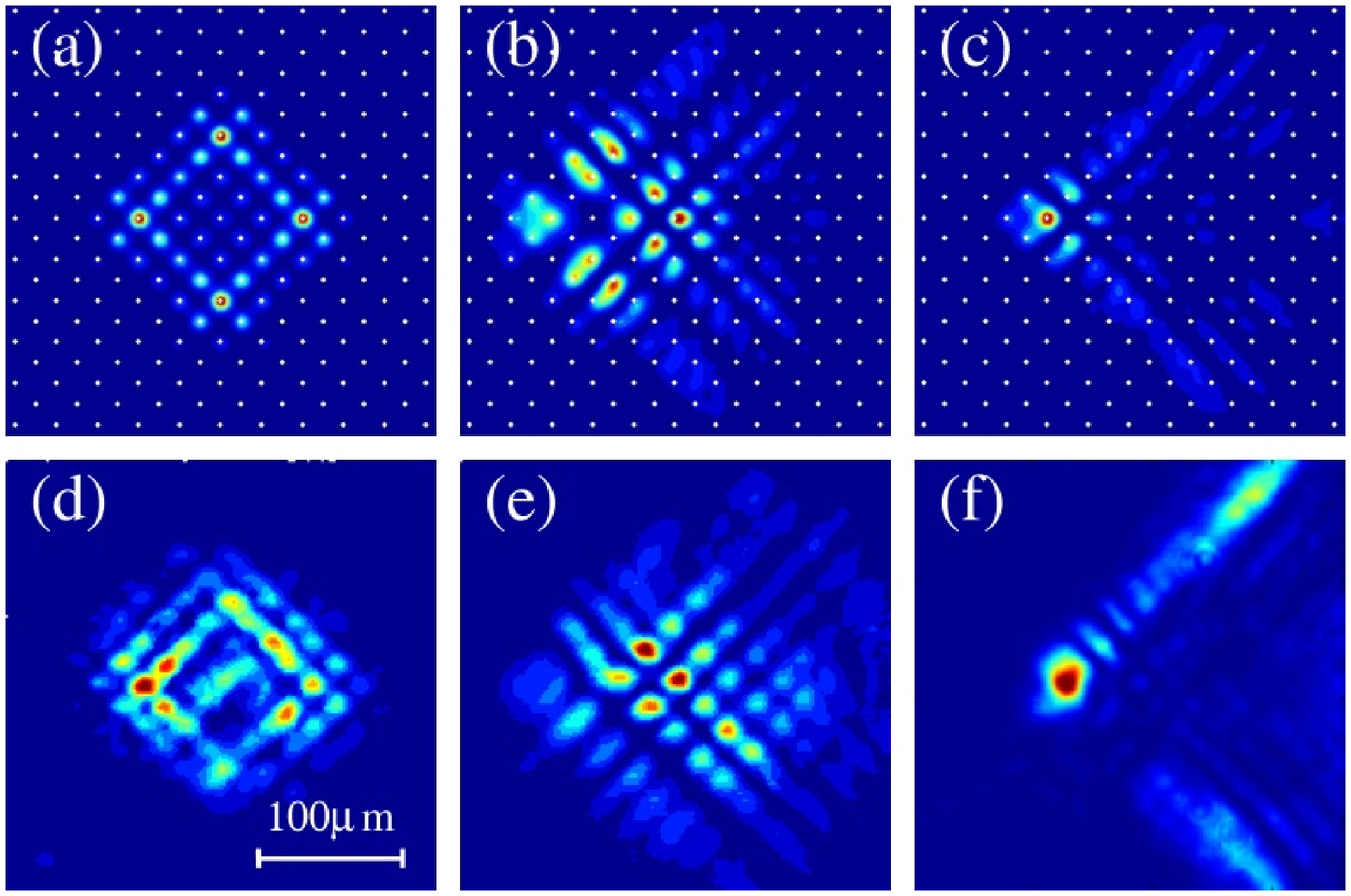}{single_site}{(color online) Light intensity profiles calculated (a-c) and measured (d-f) at the crystal output for a single site excitation. (a,d) discrete diffraction with no transverse gradient, (b,e) weak and (c,f) strong index gradients.}

It should be noted that, as we tilt the input beam, we lose information about the absolute position of the Bloch trajectory. Nevertheless, we can study the inter-band tunneling by comparing the output intensity profiles.
Additionally, fully conclusive picture of the beam evolution during a Bloch oscillation and the related Zener tunneling can be obtained by {\em monitoring the Fourier spectrum} of the output field. This is realized by placing a lens in the beam path at the position of the image plane of the back side of the crystal. Then the focal plane of the lens contain the corresponding Fourier image. In this way various stages of the tunneling process can be resolved [see Fig.~\rpict{experiment}, bottom row]. The plot in Fig.~\rpict{experiment}(a) corresponds to the initial beam tilt of $-0.095^\circ$ (negative angle corresponds to
an initial motion of the beam against the force produced by the gradient). For this angle the beam intensity profile at the output facet of the crystal is strongly modulated with adjacent maxima being approximately out-of-phase. Corresponding intensity distribution in the Fourier domain shows that the beam propagates at the Bragg angle ($0.32^\circ$) in the M-symmetry point of the lattice. Increasing the input angle allows us to scan through different positions inside the Bloch period and to monitor the light tunneling into the higher-order bands. Evidently, only one peak lies inside the first Brillouin zone for all the incident angles, and it corresponds to the field which undergoes Bloch oscillations. The other three peaks representing tunneled beams lie outside the boundaries of the first Brillouin zone, thus they belong to the second transmission band. For an angle of $-0.035^\circ$ [Fig.~\rpict{experiment}(b)] the beam just experiences its first Bragg reflection and the three tunneled beams emerge from all M-symmetry points of the lattice. At larger angles all the beams get accelerated again by the index gradient, and the central part of the beam completes a full Bloch oscillation, see Figs.~\rpict{experiment}(c-e).
We note that changing the angle between the lattice and the index gradient leads to asymmetry in scattered and tunneled beams, but the qualitative picture remains the same.

As discussed above, the Zener tunneling accompanying Bloch oscillations of a broad beam that covers several sites of the lattice occurs only when the beam approaches the edge of the Brillouin zone. A drastically different scenario has been predicted~\cite{Witthaut:2004-41:NJP} for an input beam which is initially very narrow, comparable in its size with a single site of the 2D lattice. Such a beam excites simultaneously waves with wavevectors distributed over the whole Brillouin zone. This results in a symmetric breathing of the beam in the first band as it periodically diffracts and refocuses in propagation~\cite{Pertsch:1999-4752:PRL}. Because the narrow beam excites modes with the wave vectors distributed over the whole Brillouin zone, there always exists a component which propagates in the vicinity of the band edge. Hence, the tunneling occurs now continuously over the whole Bloch period. To observe this effect we use a tightly focused probe beam ($20\mu$m) which is centered at one of the lattice sites and launched into the crystal at normal incidence. We record the light intensity distribution at the exit facet of the crystal varying the strength of the transverse index gradient. Figures~\rpict{single_site}(a,d) show the discrete diffraction of a narrow beam in the lattice with no superimposed index gradient. Note the almost perfect spatial symmetry of the field distribution which reflects the geometry of the lattice. In the case of a rather weak gradient [Figs.~\rpict{single_site}(b,e)] the light travels approximately two thirds of the Bloch period before reaching the end of the crystal. The central part of the beam starts to refocus while some light already tunnels to the second band. Although no pronounced spatial separation of the light in the first and second bands occurs at the available crystal length, the induced asymmetry of the field distribution is a clear indication that tunneling takes place.
For a strong gradient in Figs.~\rpict{single_site}(c,f) almost no breathing is visible and the light in the first band is confined to the input waveguide and its direct neighbors. 

In conclusion, we have studied the propagation of light beams in a two-dimensional optically-induced photonic lattice in the presence of a transverse ramp of the refractive index. We have observed two-dimensional photonic Bloch oscillations and Zener tunneling to distinct points of the bandgap spectrum. 

This work was produced with the assistance of the Australian Research Council under the ARC Centres of Excellence program.

\end{document}